\documentclass[prb,twocolumn,aps,showpacs]{revtex4-2}
\usepackage{graphicx}
\usepackage{amsmath}
\usepackage{physics}
\usepackage{amsfonts}
\usepackage{subfigure}
\usepackage{amssymb}
\usepackage[normalem]{ulem}
\usepackage{bm}
\usepackage[usenames,dvipsnames]{color}
\usepackage{comment}
\usepackage[colorlinks, breaklinks, 
            linkcolor=NavyBlue,
            citecolor=NavyBlue,
            urlcolor=NavyBlue]{hyperref}

\usepackage[all]{hypcap} 
\usepackage{enumitem}
\usepackage[utf8]{inputenc}
\usepackage{hyperref}

\begin{document}

\title{Nonlinear Hall responses in tunable nodal Dirac semimetals}

\author{Akash Dey}
\email{akash.dey@niser.ac.in}

\address{National Institute of Science Education and Research, Jatni, Odisha 752050, India}
\address{Homi Bhabha National Institute, Training School Complex, Anushakti Nagar, Mumbai 400094, India}


\date{\today}

\begin{abstract}
We investigate the nonlinear Hall responses in tunable two-dimensional Dirac materials. In particular, we study quantum geometry-driven second and third-order nonlinear responses in a time-reversal symmetric Dirac semimetal that can host single-node, double-node, and nodal-ring depending on the model parameters. We find that the second-order Hall (SOH) response, which originates from the Berry curvature dipole, is enhanced in the single-node semimetallic phase as compared to the double-node case when inversion symmetry is broken. In contrast, the SOH vanishes in the nodal-ring semimetal as the inversion symmetry is retained. Notably, the third-order Hall response due to Berry connection polarizability becomes much larger in the nodal-ring Dirac semimetal, especially when the Fermi energy lies near the band edge, than in the single- and double-node Dirac semimetals. The reason for this contrasting behavior is attributed to the distinct distribution of the Berry connection polarizability in the Brillouin zone.

\end{abstract}

\maketitle
	
\section{\label{sec:level1}Introduction}
In recent years, the geometry of quantum states has become crucial for understanding various unconventional electronic and optical phenomena such as photogalvanic effect, nonlinear magneto-optical responses, nonlinear Nernst effect, nonlinear Hall effect, and many  more~\cite{AHEReview,Orenstein,Morimoto,Essay,Taraphder,BRODY200119}. The Berry curvature (BC), one of the key quantities that captures the geometry of quantum states, is known to drive the anomalous Hall effect in systems that break time-reversal symmetry~\cite{Di, Haldane}. Moreover, the higher-order moments of the BC, namely the Berry curvature dipole (BCD) and quadrupole (BCQ)~\cite{AHEReview, Sodemann, Cao, Giant,Korrapati} can generate nonlinear Hall effects under different symmetry conditions. Very recently, new geometric contributions to nonlinear transport have been identified, particularly those arising from interband coherence and higher-order geometric corrections. One such quantity is the Berry connection polarizability (BCP), also referred to as the band-normalized quantum metric~\cite{sala2024quantum}. The BCP measures how the interband Berry connection responds to changes in energy within the band structure, providing an intrinsic mechanism for nonlinear Hall effects beyond the conventional dipole picture~\cite{Gao, SOHE_LIU, Liu}. In addition, BCP is shown to serve as a probe of the Néel vector orientation~\cite{Chong, SOHE_LIU} even in $PT$-symmetric antiferromagnetic systems.

Dirac semimetals (DSMs) with symmetry-protected band crossings provide an excellent platform to study the effect of quantum geometry on various physical properties. Based on the dimensionality and nature of the band crossings, semimetals can be classified as Dirac~\cite{Kane,Sergey,Lizhou,Debasmita}, Weyl~\cite{soluyanov,Ling,Wan,lu2024,Roy_2022}, and nodal-ring semimetals (NRSMs)~\cite{Fang,Xue,CHANG2025,Hu_2025,Ming-Jian,Lei,Vivek,pandey2025disorder}. While DSMs and Weyl semimetals (WSMs) feature discrete band-touching points, NRSMs exhibit one-dimensional crossings that form continuous loops in momentum space~\cite{Armitage,Chiu}. Breaking certain symmetries can induce transitions between these phases. For example, a DSM can transform into a WSM when either time-reversal or inversion symmetry is broken~\cite{Elena,Jing}. In some ferromagnetic materials such as $\mathrm{HgCr_2Se_4}$, $\mathrm{Co_2TiX}$ (X = Si, Ge, Sn), and $\mathrm{Fe_2MnX}$ (X = P, As, Sb), tuning material parameters enables transitions between NRSM and WSM phases~\cite{Gang,chang2016room,Noky,Jing}. Beyond 3D systems, two-dimensional analogs of Dirac, Weyl, and nodal-ring semimetals have been realized in transition-metal dichalcogenides, graphene heterostructures, and mixed honeycomb–kagomé lattices~\cite{Kane,ezawa2019second,chen2018,Lu_2017}. 

The presence of distinct nodal structures in materials is found to host distinct physical properties such as optical response\cite{basov_2019,PhysRevB.102.035138}, magneto-optical transition\cite{wfzd-7r9n}, chiral anomaly induced negative magnetoresistance\cite{Long,Balduini}, drumhead surface states, and unique electromagnetic responses\cite{Po-Yao}. Motivated by the distinct properties of distinct nodal structures, we aim to understand how the geometric quantities change across these nodal points and subsequently affect the nonlinear transport properties of nodal materials. To address this, we consider a low-energy model Hamiltonian of a Dirac semimetal that hosts distinct nodal structures depending on the model parameters. We then investigate how quantum geometric quantities evolve across distinct nodal phases.  A family of two-dimensional semimetals, MX with M = Pd, Pt, and X =S, Se, Te~\cite{Jin,Barati}, provides a promising platform to explore these effects. We find that BCD is significantly enhanced in the gapped single-node phase as compared to the double-node phase. This behavior arises due to the distinct distribution of the Berry curvatures near the nodal structure. As a result, the second-order Hall (SOH) response is larger in the single-node phase than in the double-node phase. In contrast, the SOH response vanishes in the nodal-ring phase because the Berry curvature is identically zero.
However, all three phases exhibit a non-zero third-order Hall (TOH) response due to the presence of a nonzero BCP. Interestingly, the nodal-ring phase shows a much stronger BCP along the ring compared to the other two phases. Consequently, we observe an enhanced TOH response near the band edge in the nodal-ring phase, compared to the band structures with single or double nodes. This behavior can be traced back to the way BCP tensors are distributed in the Brillouin zone for the different nodal configurations. Overall, these findings provide a useful pathway to identify materials with enhanced nonlinear responses among nodal systems.

The paper is organised as follows: In Sec.~\ref{Model Hamiltonian}, we introduce the model Hamiltonian and the underlying symmetries. This is followed by a formal derivation for expressions for the electrically driven higher order Hall conductivities in terms of quantum geometric quantities in Sec.~\ref{Theoretical Background}. We then present numerical results showing the evolution of second-order and third-order Hall responses across various nodal phases in Section \ref{BCD induced} and \ref{BCP induced}. Finally, in Sec.~\ref{Summary}, we summarize our main findings and discuss the future outlook.

\section{\label{Model Hamiltonian}Model Hamiltonian}
\begin{figure}[htbp]
	\centering
	\includegraphics[width=1\linewidth]{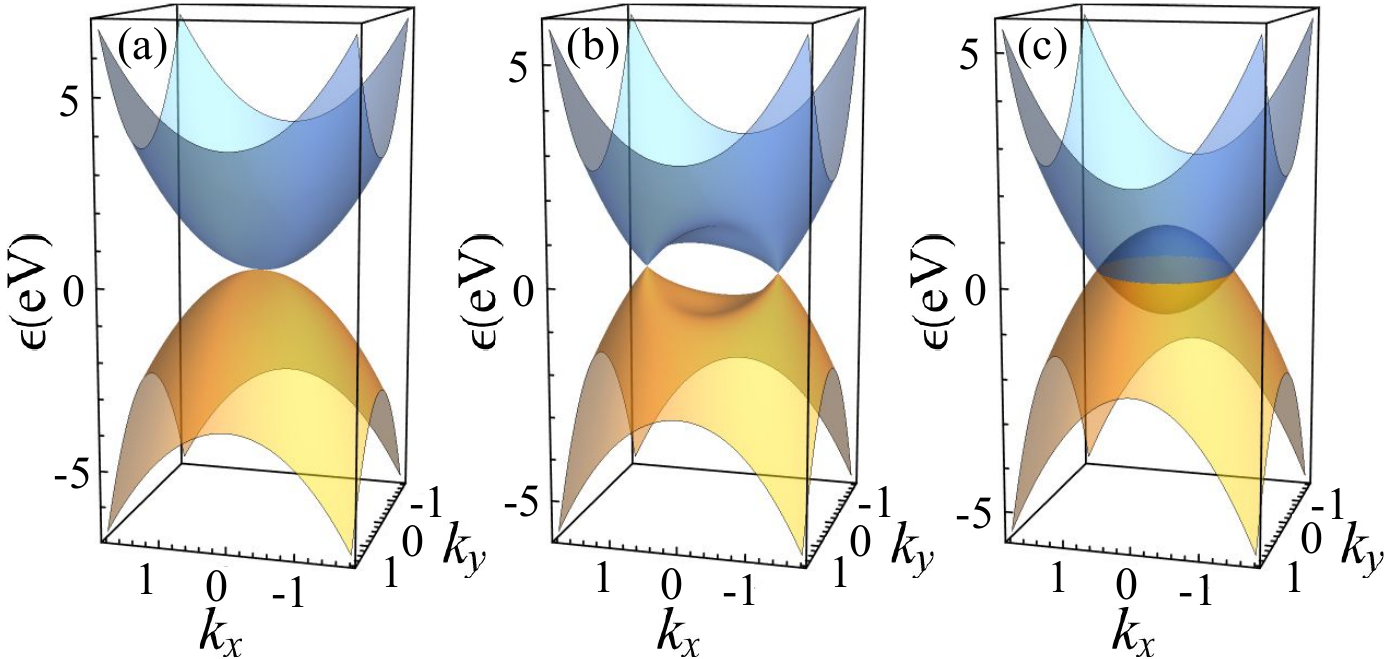}
        \caption {Energy spectrum of Eq.~\ref{Hamiltonian} for different values of parameters $\gamma$ and $k_0$, keeping $\lambda$ fixed. (a) SN phase with single nodal point at  $\mathbf{k} = 0$ for $k_0=0$ and $\gamma=1.0~\mathrm{eV\!\cdot\!\mathring{A}}$, (b) DN phase with two nodal points located at $\mathbf{k} = (\pm k_0,0)$ for $k_0=1.0~\mathrm{\mathring{A}^{-1}}$ and $\gamma=1~\mathrm{eV\!\cdot\!\mathring{A}}$, and (c) NR phase with a nodal ring of radius $k_0=1.0~\mathrm{\mathring{A}^{-1}}$ and $\gamma=0$. Here we take $\lambda=1.0~\mathrm{eV\!\cdot\!\mathring{A}^2}$ and momentum $k_x$, $k_y$ are in units of $\text{\AA}^{-1}$.}
	\label{dispersion}
\end{figure}

We consider a low-energy model Hamiltonian for a two-dimensional semi-metallic system\cite{Jin,Rahimpoor,Pandey,Po-Yao} as 
\begin{equation}
    H= \bold{d}(\bold{k})\cdot\boldsymbol{\sigma},
    \label{Hamiltonian}
\end{equation}
where $\boldsymbol{\sigma} = (\sigma_x, \sigma_y, \sigma_z)$ are the Pauli matrices acting in the pseudospin space, $\mathbf{d}(\mathbf{k}) = (d_x, d_y, d_z) =(\lambda(k^2 - k_0^2), \gamma k_y, 0)$ with $\mathbf{k} = (k_x, k_y)$, denoting the crystal momentum. Here, $\lambda$ represents the inverse of the band mass, $\gamma$ characterizes the Fermi velocity along the $y$ direction and $k_0$ is the parameter that determines the position of the nodal points. The corresponding energy dispersion is given by
\begin{equation}
    \epsilon_{\pm}=\pm\sqrt{(\lambda(k^2-k_0^2))^2+(\gamma k_y)^2},
    \label{spectrum}
\end{equation}
where $\pm$ denote the conduction and valence bands, respectively. By tuning the parameters $\gamma$ and $k_0$ with $\lambda$ fixed, the system can be continuously driven through distinct gapless semimetallic phases, as illustrated in Fig.~\ref{dispersion}. 

For $k_0 = 0$ and $\gamma\ne 0$, the two energy bands touch at a single point at $\mathbf{k} = 0$, corresponding to a single-node (SN) semimetallic phase. For both $k_0\ne0$ and $\gamma\ne0$, the node splits into two distinct band-touching points at $\mathbf{k} = (\pm k_0,0)$, leading to a double-node (DN) phase. Conversely, when $\gamma = 0$ and $k_0\ne0$, the discrete nodes expand into a continuous closed loop in momentum space, forming a nodal ring (NR) semimetal with a ring radius equal to $k_0$. When a finite gap parameter $\Delta$ is introduced ($d_z=\Delta$), the system no longer exhibits semimetallic behavior. Nevertheless, for small values of $\Delta$, the energy dispersion continues to reflect the underlying nodal character, with the energy bands extrema occurring near $k=0$ and $k=k_0$~\cite{Barati}.
Note that for SN and DN phases, the system is invariant under mirror symmetry $M_x$ and the discrete nodal points are located on the mirror line along $x$ direction. In contrast, in the NR phase, the system possesses rotational symmetry. 
Note that time reversal symmetry (TRS) with $\Theta=\mathcal{K}$, where $\mathcal{K}$ denotes the complex conjugation operator, is preserved in all three cases. However, the inversion symmetry (IS) is broken only in SN and DN phases but remains preserved in NR phase.

\section{\label{Theoretical Background} Theoretical Background}
We start with a general formalism to evaluate different order Hall conductivity for systems with time-reversal symmetry. Using the Boltzmann transport formalism, the total electric current density is found to be
\begin{equation}
    \bold{j}=-e\int_k f(\bold{k}) \bold{\Dot{r}}(\bold{k}),
    \label{ja}
\end{equation}
where $e$ is the electron charge, $\int_k$ = $\int d^dk/(2\pi)^d$, $\bold{r}$ is the position vector of the wavepacket, and $f$ denotes the non-equilibrium distribution function. Under the influence of an applied electric field $\boldsymbol{E}(t)$ = Re[$\bold{E_0}e^{i\omega t}$], the perturbation to the system is described by $\mathrm{H^{\prime}}$ = $ e\bold{E.r}$, which induces a position shift in the wave packet~\cite{Gao}. The semiclassical equations of motion, including higher-order corrections due to the electric field, can be expressed as (set $e=\hbar=1$)~\cite{Gao2015,Di}
\begin{equation}
    \bold{\Dot{r}}(\bold{k})=\bold{\nabla}_{\bold{k}} \epsilon^c(\bold{k})-\mathbf{E}\times \bold{\Omega}^c(\bold{k})
    \label{rdot},
\end{equation}
with $\bold{\Dot{k}}=-\bold{E}$, $\epsilon^c$ and $\bold{\Omega}^c$ are the higher-order corrections to band energy and Berry curvature, respectively. They can be expressed as 
\begin{equation}
    \epsilon^c(\bold{k})=\sum_{i=0}^2\epsilon(\bold{k})^{(i)},~~~~~ \bold{\Omega}^c(\bold{k}) = \bold{\nabla}_\bold{k} \times \sum_{i=0}^1 \bold{A}^{(i)}(\bold{k}),
\end{equation}
where $\bold{A}^{(i)}$ is the $i-$th order correction to the Berry connection. For the zeroth order correction in energy and BC, the Hall current density derived from equation (\ref{ja}) is given by ~\cite{BANDYOPADHYAY2024100101, Sodemann}:
\begin{equation}
j_{\alpha}=\sigma_{\alpha\beta}E_{\beta}+\zeta_{\alpha\beta\gamma}E_{\beta}E_{\gamma}.
\end{equation}
The linear and second-order Hall conductivity $\sigma_{\alpha\beta}$ and $\zeta_{\alpha\beta\gamma}$ respectively reads off
\begin{align}
    \sigma_{\alpha\beta}&= -\varepsilon_{\alpha\beta\gamma}\int_k \Omega_{\gamma}(\bold{k}) f_{eq}(\bold{k}),
    \label{Linear}\\
    \zeta_{\alpha\beta\gamma}&=-\varepsilon_{\alpha\gamma\delta}\frac{\tau}{2(1+i\omega\tau)}D_{\beta\delta}.
    \label{SOHE}
\end{align}
Here $(\alpha,\beta,\gamma,\delta)$ $\in$ $(x,y,z)$ are applied field directions, $f_{eq}$ is the equilibrium distribution function, $\tau$ is the scattering time, and $\omega$ is the driving frequency. The quantity $D_{\beta\delta}$ in equation~(\ref{SOHE}) is known as Berry curvature dipole\cite{Sodemann} and it is defined as 
\begin{equation}
    D_{\beta\delta}=\int_k f_{eq}(\bold{k})(\partial_{k_{\beta}}\Omega_{\delta}(\bold{k}))= -\int_k v_{\beta}(\bold{k})\Omega_{\delta}(\bold{k})\frac{\partial f_{eq}}{\partial \epsilon^{(0)}},
    \label{SOHEBCD}
\end{equation}
where $\partial_{k_{\beta}}=\frac{\partial}{\partial k_{\beta}}$, $v_{\beta}=\partial \epsilon^{(0)}/\partial k_{\beta}$ is the band velocity and the factor $\partial f_{eq}/\partial \epsilon^{(0)}$ restricts the integral to states near the Fermi surface at low temperature.

To capture the third-order Hall effect, we include the second-order correction to the band energy because the first-order term is gauge-dependent and vanishes in the wave packet picture~\cite{Gao,Tanay}. We also incorporate the first-order correction to the Berry curvature. In the presence of the perturbation $\mathrm{H^{\prime}}$, the first-order correction to the Bloch wavefunction is
\begin{equation}
    \ket{u_{m}^{(1)}(\bold{k})}=\sum_{n\neq m}\frac{e \bold{E}.\bold{A}_{nm}^{(0)}(\bold{k})\ket{u_{n}^{(0)}(\bold{k})}}{\epsilon_{m}^{(0)}(\bold{k})-\epsilon_{n}^{(0)}(\bold{k})}, 
\end{equation}
where $\bold{A}_{nm}^{(0)}(\bold{k})=\bra{u_{m}^{(0)}(\bold{k})} i\bold{\nabla_k}\ket{u_{n}^{(0)}(\bold{k})}$ is the zeroth-order interband Berry connection and $m,n$ are the band indices. Then the first order correction in BC is $\bold{\Omega}^{(1)}(\bold{k})=\bold{\nabla_k}\times \bold{A}^{(1)}(\bold{k})$, where $\bold{A}^{(1)}$is 

\begin{equation}
A^{(1)}_{m,\alpha}(\bold{k})=2Re\sum_{n\neq m}\frac{A^{(0)}_{m n,\alpha}(\bold{k})A^{(0)}_{n m,\beta}(\bold{k})}{\epsilon_{m}^{(0)}(\bold{k})-\epsilon_{n}^{(0)}(\bold{k})}E_b=G_{m,\alpha\beta}E_\beta.
  \label{A1BCP}
\end{equation}
Here, we introduce the second-rank tensor
\begin{equation}
   G_{m,\alpha\beta}=2~\mathrm{Re}\sum_{n\neq m}\frac{A^{(0)}_{m n,\alpha}(\bold{k})A^{(0)}_{n m,\beta}(\bold{k})}{\varepsilon_{m}^{(0)}(\bold{k})-\varepsilon_{n}^{(0)}(\bold{k})},
   \label{eq:BCP}
\end{equation}
referred to as the \textit{Berry connection polarizability} tensor, represents another key geometric quantity of the system. Moreover, the second-order correction to the band energy can be expressed in terms of this BCP tensor as follows:
\begin{equation}
    \epsilon_{m}^{(2)}(\bold{k})=\sum_{n\neq m}\frac{|\bra{u_{m}^{(0)}(\bold{k})}H^{\prime}\ket{u_{n}^{(0)}(\bold{k})}|^2}{\epsilon_{m}^{(0)}(\bold{k})-\epsilon_{n}^{(0)}(\bold{k})}=\frac{1}{2}G_{\alpha\beta}E_{\alpha}E_{\beta}.
\end{equation}
Next, we proceed to determine the non-equilibrium distribution function by solving the semiclassical Boltzmann equation under relaxation time approximation~\cite{ashcroft1976solid}, which yields
\begin{equation} 
    \bold{\Dot{k}}.\bold{\nabla_k}f(\bold{k})=\frac{f_{eq}(\bold{k})-f(\bold{k})}{\tau}.
    \label{eq:BZrelax}
\end{equation}
For the solution of equation~(\ref{eq:BZrelax}), we consider the following ansatz:
\begin{equation}
    f(\bold{k})=\sum_{i=0}^{\infty}(\tau \bold{E}.\bold{\nabla_k})^i f_{eq}(\epsilon^c(\bold{k})).
     \label{fansatz}
\end{equation}
By substituting Eqs.~(\ref{rdot}) and (\ref{fansatz}) in equation~(\ref{ja}), we obtain the third-order current density as (see appendices for details) \cite{Liu,Ojasvi,Tanay} 
\begin{widetext}
\begin{align}
    j^{(3)}&=\int_k(\bold{E}\times\bold{\Omega}^{(0)}(\bold{k}))\epsilon^{(2)}(\bold{k})f_{eq}^{\prime}(\bold{k})-
    \tau\int_k\bold{\nabla_k}\epsilon^{(0)}(\bold{k})(\bold{E.\nabla_k})\epsilon^{(2)}(\bold{k})f_{eq}^{\prime}(\bold{k})-\tau\int_k \bold{\nabla_k}\epsilon^{(2)}(\bold{k})(\bold{E.\nabla_k})f_{eq}(\bold{k}) \nonumber\\&+\tau\int_k(\bold{E}\times\bold{\Omega}^{(1)}(\bold{k}))(\bold{E.\nabla_k})f_{eq}(\bold{k})
    +\tau^2\int_k(\bold{E}\times\bold{\Omega}^{(0)}(\bold{k}))(\bold{E.\nabla_k})^2f_{eq}(\bold{k})
    -\tau^3\int_k \bold{\nabla_k}\epsilon^{(0)}(\bold{k})(\bold{E.\nabla_k})^3f_{eq}(\bold{k}).
    \label{TOHcurrentdensity}
\end{align}    
\end{widetext}
Since both the $\tau$-independent and $\tau^2$ terms in the right-hand side of equation~\ref{TOHcurrentdensity} are odd under time-reversal symmetry, neither contributes to the current density. Therefore, only the $\tau$ and $\tau^3$ terms contribute to the third-order current and each term has a clear physical origin. The second term comes from the second-order energy correction inside the distribution function. The third term follows from the second-order field correction to the band velocity. The fourth term comes from the anomalous velocity produced by the first-order field correction to the Berry curvature. The final term, which is proportional to $\tau^3$, originates from the band velocity and gradient of the distribution function. Since we are interested in the BCP induced TOH current, we will only consider the terms proportional to $\tau$. Finally, we can write the TOH current density as $j_{\alpha}^{(3)}=\chi_{\alpha\beta\gamma\delta}E_{\beta}E_{\gamma}E_{\delta}$, where the third order conductivity linear in $\tau$ upon simplification is given by (see appendices for details)
\begin{align}
    \chi_{\alpha\beta\gamma\delta}&= \tau \int_k (\partial_{k_{\alpha}}\partial_{k_{\beta}} G_{\gamma\delta}-\partial_{k_{\alpha}}\partial_{k_{\delta}} G_{\beta\gamma}+\partial_{k_{\beta}}\partial_{k_{\delta}} G_{\alpha\gamma})f_{eq}(\bold{k})\nonumber\\&-\frac{\tau}{2}\int_k v_{\alpha}(\bold{k}) v_{\beta} (\bold{k})G_{\gamma\delta}f_{eq}^{\prime\prime}(\bold{k})
    \label{TOHE Conductivity}.
\end{align}

\section{Results}
The transition between the gapped nodal phases leads to changes in the underlying geometric quantities, including the BC, BCD, and BCP. These changes directly influence the resulting nonlinear Hall responses. To illustrate these connections, we begin by analyzing the BCD and its role in the second-order Hall effect, followed by a discussion of the BCP and the resulting third-order Hall response.
\subsection{\label{BCD induced}BCD induced Second order Hall effect.} The generic form of the Hamiltonian in equation~\ref{Hamiltonian} allow us to easily evaluate the Berry curvature as (see appendices for details)
\begin{align}
    \Omega_z^{\pm}&=\frac{\mathbf{d}(\mathbf{k}) \cdot 
\left( \partial_{k_x} \mathbf{d}(\mathbf{k}) 
\times 
\partial_{k_y} \mathbf{d}(\mathbf{k}) \right)}
{2 |\mathbf{d}(\mathbf{k})|^3}\nonumber\\&=\pm \frac{\lambda k_x \gamma \, \Delta}{\left[(k^2-k_0^2)^2 \lambda^2 + k_y^2 \gamma^2 + \Delta^2\right]^{3/2}}.
\label{Berry curvature}
\end{align}
Evidently, $\Omega_z$ is an odd function of $k_x$. Consequently, the total BC integrated over the Brillouin zone vanishes (equation~\ref{Linear}), ensuring that the linear Hall conductivity is zero in this time-reversal-invariant system. We therefore compute the second-order conductivity arising from the BCD.  

The velocity components $v_x^{\pm} = 2k_xd_x/\epsilon_{\pm}$ and $v_y^{\pm} = k_y(2d_x+\gamma^2)/\epsilon_{\pm}$, both are odd function in $k_x$ and $k_y$ respectively. Thus the integrand $v_x\Omega_z$ for the $x$ component of BCD  equation~\ref{SOHEBCD}, becomes an even function under momentum inversion and contributes a finite value of $D_{xz}$. 
In contrast, for the $y$ component, the integrand $v_y\Omega_z$ becomes an odd function that integrates to zero, thereby suppressing the $D_{yz}$. This is consistent with the underlying mirror symmetry $M_x$, which forbids any BCD response along $y$-direction~\cite{Sodemann,Awadhesh}. Importantly, this finite BCD emerges even without any explicit {\it tilt} in the band structure. It arises inherently from the distribution of the Berry curvature in momentum space combined with the velocity anisotropy, demonstrating that nonlinear Hall responses can exist in untilted systems when the underlying band geometry lacks complete inversion symmetry in momentum space. The magnitude and profile of this response further depend on the nodal configuration of the system as will be discussed next. 

\begin{figure}[htbp]
	\centering
	\includegraphics[width=1.0\linewidth]{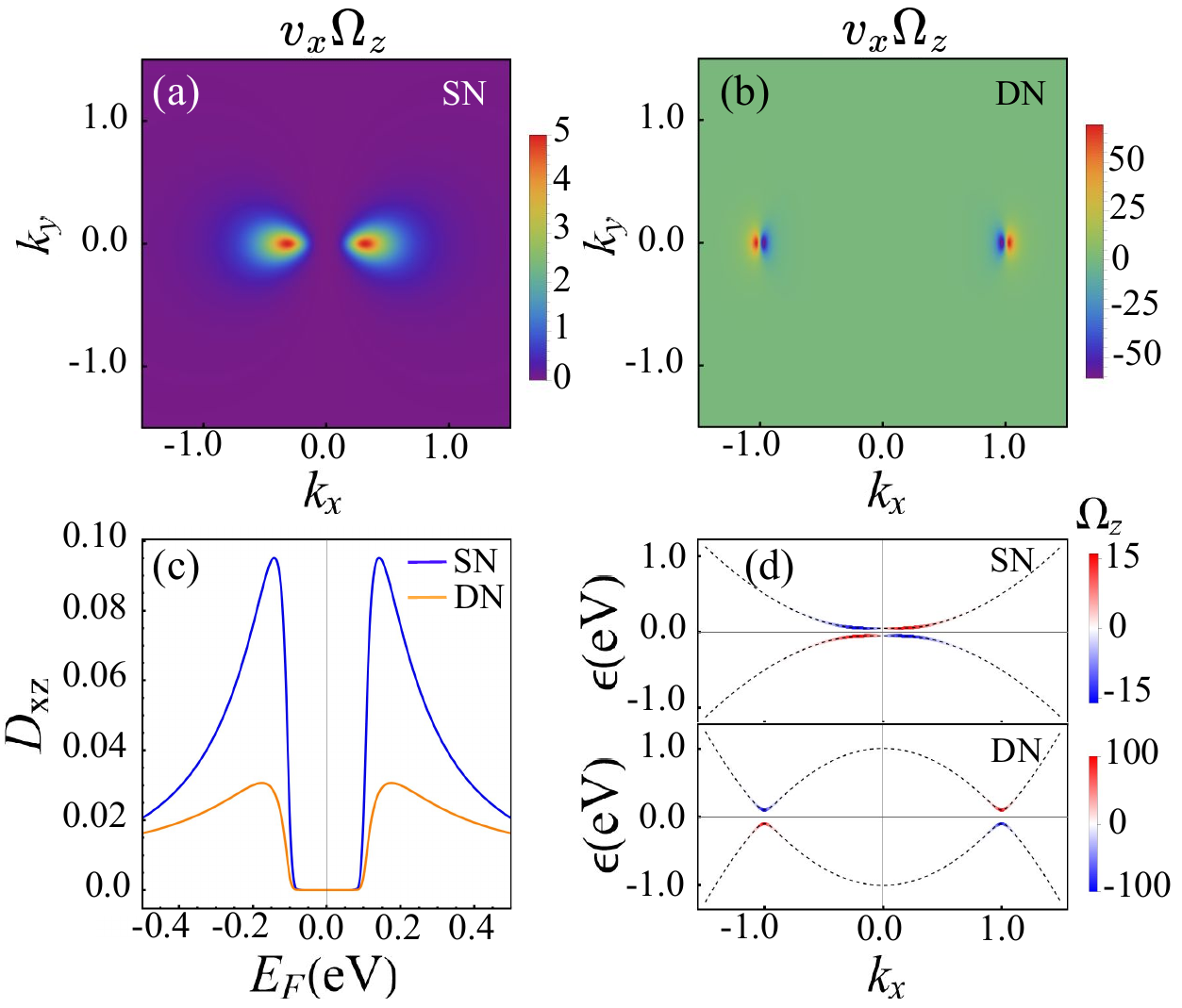}
        \caption {(a,b) Density plots showing the Berry curvature ($\Omega_z$) multiplied by the velocity component for the single-node and double-node phases, respectively. (c) Plot of the Berry curvature dipole $D_{xz}$ for both these phases. Here, we take the gap parameter  $\Delta=0.1~\mathrm{eV}$. (d) Density plot of Berry curvature along the energy spectrum in the $k_y=0$ plane for both the SN and DN phases.  Here $\Omega_z$ and $D_{xz}$ are in units of $\text{\AA}^2$ and $\text{\AA}$ respectively.}
	\label{SOHE Plots}
\end{figure}

For the NR phase with $\gamma=0$, the BC vanishes, and therefore this phase possesses no second-order response. We therefore focus only on the finite responses that arise in the SN and DN phases. In the SN phase, Fig.~\ref{SOHE Plots}(a) shows the density plot of the integrand $v_x\Omega_z$. Clearly, this quantity is strongly concentrated near the origin and displays two symmetric positive regions due to $k_x^2$ term in the numerator, leading to a large net dipole,  as seen in Fig.~\ref{SOHE Plots}c. In contrast, in the DN phase, the integrand shown in Fig.~ (\ref{SOHE Plots}(b)) is concentrated around $k_x=\pm k_0$, forming a dipole-like structure with opposite signs. As a result, the net dipole contribution is significantly reduced as illustrated in Fig.~\ref{SOHE Plots}(c). In both cases, the response vanishes in the gap region as the BCD is a Fermi surface property. Also, the response reaches to its maximum value near the band edge as BC becomes maximum at the band edges (see Fig.~\ref{SOHE Plots}(d)). Note that as the Hamiltonian is TRS invariant, the BCP induced intrinsic second-order Hall response is absent~\cite{Das} in all these cases.

\subsection{\label{BCP induced}BCP induced third order Hall effect:}
Since the third-order Hall response is not restricted by TRS and IS~\cite{Liu,Tanay}, we observe a finite BCP induced TOH response in all three phases. Using equation~\ref{eq:BCP},  different components of the BCP tensor for the Hamiltonian in equation~\ref{Hamiltonian} are found to be (see appendices for details)
 \begin{align}
     G_{xx}^{\pm}&=\pm \frac{k_x^2 \left( k_y^2 \gamma^2 + \Delta^2 \right) \lambda^2}
{\left[(k^2-k_0^2)^2 \lambda^2 + k_y^2 \gamma^2 + \Delta^2\right]^{5/2}},\nonumber\\
     G_{yy}^{\pm}&=\pm\frac{\gamma^2 (\Delta^2 + \left( k_0^2 - k_x^2 + k_y^2 \right)^2 \lambda^2 )+4k_y^2 \Delta^2 \lambda^2}
{4\left[(k^2-k_0^2)^2 \lambda^2 + k_y^2 \gamma^2 + \Delta^2\right]^{5/2}}
,\nonumber\\
     G_{xy}^{\pm}&= G_{yx}^{\pm}=\pm\frac{k_x k_y (\left( k_0^2 - k_x^2 + k_y^2 \right) \gamma^2 + 2 \Delta^2)\lambda^2}
{2\left[(k^2-k_0^2)^2 \lambda^2 + k_y^2 \gamma^2 + \Delta^2\right]^{5/2}}.
\label{BCP components}
 \end{align}

\begin{figure}[htbp]
	\centering
	\includegraphics[width=1\linewidth]{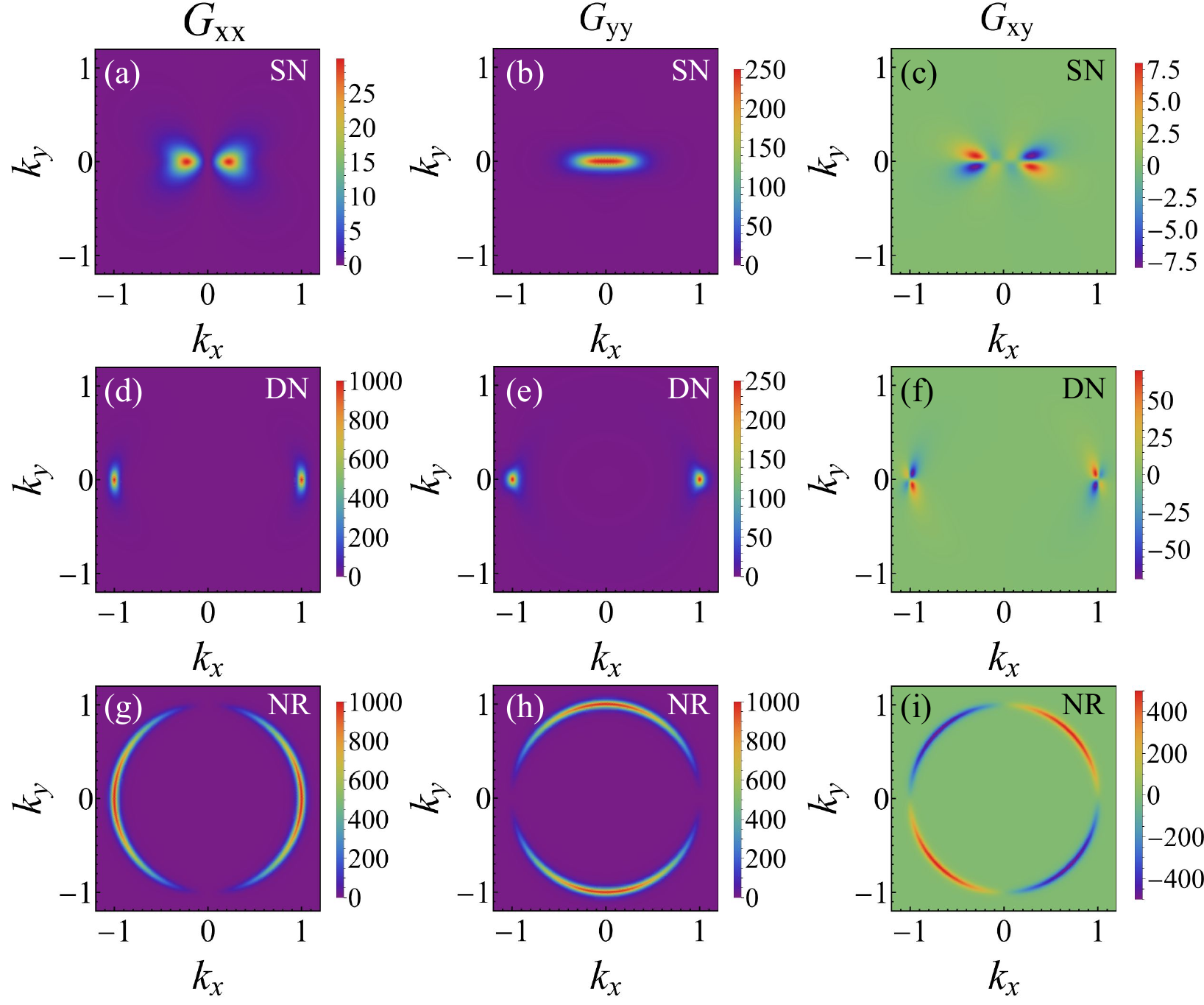}
	\caption{Density plots of the BCP tensor components $G_{xx}$(left column), $G_{yy}$(middle column) and $G_{xy}$(middle column)  for all three phases, obtained from equation~\ref{BCP components}. The top row (a–c), middle row (d–f) and bottom row (g–i) correspond to the SN, DN and NR phases, respectively. The plots are shown only for the conduction band. Here we consider the gap $\Delta=0.1~\mathrm{eV}$.}
    \label{BCP tensor densityplot}
\end{figure} 
In Fig.~\ref{BCP tensor densityplot}, we have shown the density plots of these BCP tensor components across different nodal phases. Each phase develops its own characteristic BCP structure, and these differences strongly influence the corresponding third-order Hall responses. In the SN phase, one might expect a monopole-like structure near $\boldsymbol{k}=0$ for the diagonal components, similar to that of a single Dirac-point system~\cite{Liu}. However, in the present model, the $G_{xx}$ component of the BCP displays two symmetric lobes along the $k_x$-axis, as shown in Fig.~\ref{BCP tensor densityplot}(a). As before, this feature originates from the $k_x^2$ factor appearing in the numerator of $G_{xx}$ in equation~\ref{BCP components}. This in turn forces the contribution to vanish exactly at the origin. However, in the immediate vicinity of the origin, the denominator becomes small, enhancing the response, leading to two symmetric maxima near $k_x$=0. In contrast, the $G_{yy}$ and $G_{xy}$ exhibit a monopole and a quadrupole-like structure as illustrated in Fig.~\ref{BCP tensor densityplot}(b) and \ref{BCP tensor densityplot}(c) respectively, similar to a single Dirac-point system. Note that $G_{yy}$ is significantly larger than both $G_{xx}$ and $G_{xy}$ due to constant numerator at the origin.

In the DN phase, both the diagonal components $G_{xx}$, $G_{yy}$ exhibit two localized monopoles, as shown in Fig.~\ref{BCP tensor densityplot}(d) and ~\ref{BCP tensor densityplot}(e) respectively. The off-diagonal term $G_{xy}$ shows two quadrupole-like patterns located at $k_x=\pm k_0$, as illustrated in Fig.~\ref{BCP tensor densityplot}(f). 
Notably, following equation~\ref{BCP components}, the $G_{xx}$ and $G_{xy}$ components exhibit stronger peak intensities as compared to the SN phase. This is attributed to a finite numerator at nodes $k_x=\pm k_0$ as compared to the SN case with $k=0$. In contrast, the $G_{yy}$ component shows identical peak intensity proportional to $\gamma^2/4\Delta^3$ near ($k_y\approx 0$, $k_x\approx\pm k_0$) and ($\boldsymbol{k}\approx0$) for DN and SN phase respectively. 

In contrast to both the SN and DN phases, the NR phase shows a ring-like BCP pattern, reflecting the underlying nodal ring geometry, as shown in Fig~\ref{BCP tensor densityplot}(g-i). The $G_{xx}$ and $G_{yy}$ components contain $k_x^2$ and  $k_y^2$ in the numerator, which results in maximum density along $k_x$ and $k_y$ direction and forms two circular ridges following the ring, as illustrated in Fig~\ref{BCP tensor densityplot}(g) and Fig~\ref{BCP tensor densityplot}(h) respectively. Conversely, the $G_{xy}$ component is proportional to $k_xk_y$, which leads to alternating density extrema in adjacent quadrants and vanishes along $k_x=0$, $k_y=0$ direction, shown in Fig~\ref{BCP tensor densityplot}(i). The overall magnitude of all components is larger in the NR phase because of the absence of $\gamma k_y\sigma_y$ term in the denominator, which effectively amplifies their intensity. For the $G_{xx}$ component in both NR and DN phases, the dominant contribution arises near $k_y\approx 0$, and $k_x=\pm k_0$, where the peak magnitude is proportional to $k_0^2/\Delta^3$. Hence, the $G_{xx}$ component exhibits nearly identical peak intensities in both the NR and DN phases, even though the underlying geometric structures are different.  

\begin{figure}[htbp]
	\centering
	\includegraphics[width=1.0\linewidth]{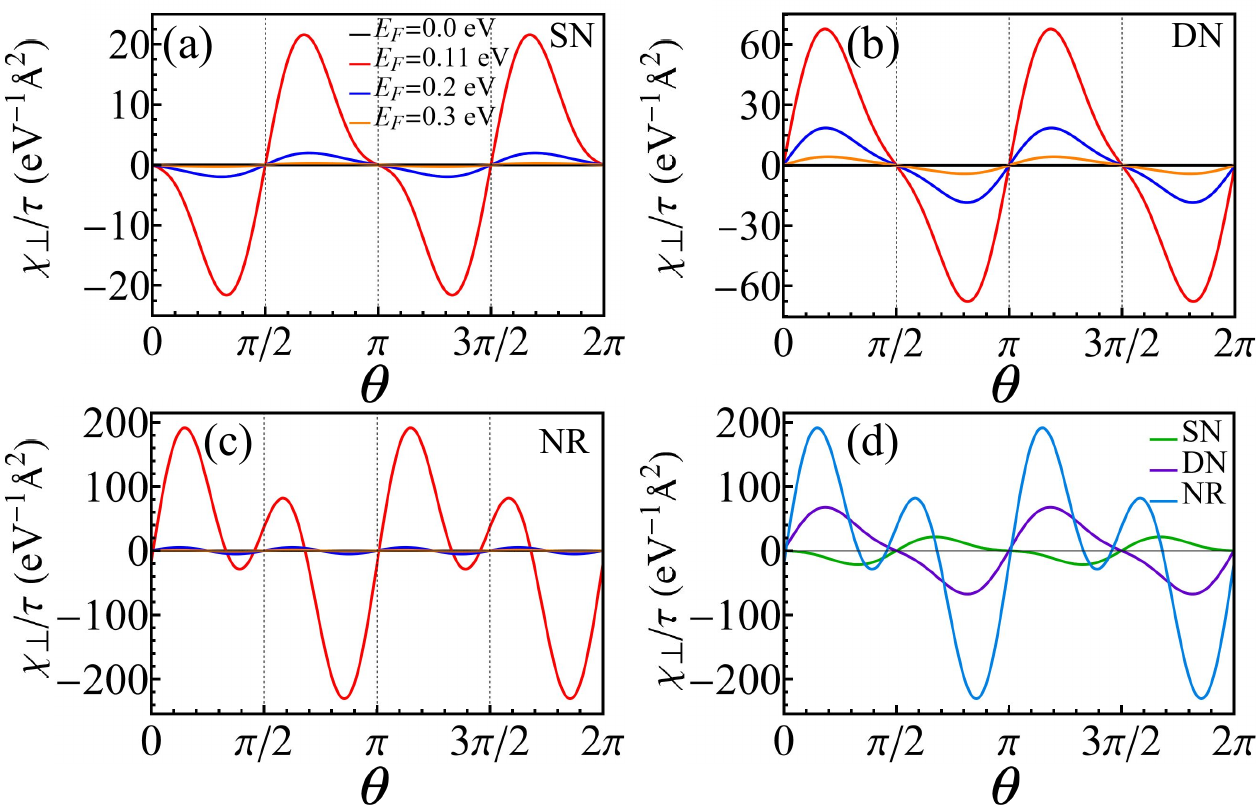}
	\caption{The transverse TOH conductivity for different nodal phases. (a-c) shows the variation of the transverse TOH response $\chi_{\perp}$ as a function of the applied electric field direction $\theta$ for different Fermi energies $E_F$ in the SN, DN and NR phase respectively. (d) presents the maximum transverse TOH responses together in different phases near the band edge at $E_F=0.11~\mathrm{eV}$. We take temperature $T$=50K.} 
    \label{TOHE plots}
\end{figure}

Using the obtained BCP tensor components, we now numerically compute the corresponding third-order conductivity tensor as defined in equation~\ref{TOHE Conductivity}. In nonlinear Hall experiments, one measures the transverse current response using an in-plane electric field $\bold{E}=(E \cos{\theta}, E \sin{\theta})$ oriented at an angle $\theta$ to the x-axis. Focusing on the transverse third-order Hall conductivity $\chi_{\perp}(\theta)=j^{(3)}_{\perp}/E^3$, we derive its explicit dependence on the applied electric field direction $\theta$  and obtain 
\begin{widetext}
\begin{align}
    \chi_{\perp}(\theta)&= (3 \chi_{21}-\chi_{11})\sin{\theta}\cos^3{\theta}+3 (\chi_{31}-\chi_{13})\sin^2{\theta}\cos^2{\theta}+(\chi_{22}-3\chi_{12})\sin^3{\theta}\cos{\theta}-\chi_{14}\sin^4{\theta}+\chi_{41}\cos^4{\theta},
\end{align}
\end{widetext}
where $\chi_{11}=\chi_{xxxx}$,  $\chi_{12}=(\chi_{xxyy}+\chi_{xyxy}+\chi_{xyyx})/3$, $\chi_{13}=(\chi_{xxxy}+\chi_{xxyx}+\chi_{xyxx})/3$, $\chi_{14}=\chi_{xyyy}$, $\chi_{22}=\chi_{yyyy}$, $\chi_{41}=\chi_{yxxx}$, $\chi_{31}=(\chi_{yyyx}+\chi_{yyxy}+\chi_{yxyy})/3$, and $\chi_{21}=(\chi_{yyxx}+\chi_{yxyx}+\chi_{yxxy})/3$.

Note that the SN and DN phases preserve the $M_x$ mirror symmetry. This symmetry ensures that the tensor components involving an odd number of $x$ and $y$ in the index (like $\chi_{xyyy}$, $\chi_{yxxx}$) must vanish~\cite{Liu,Tanay}. As a result, the tranverse TOH conductivity $\chi_{\perp}$ vanishes when $\theta$ becomes integer multiples of $\pi/2$, i.e., along and perpendicular to the mirror line in Fig.~\ref{TOHE plots}(a) and Fig.~\ref{TOHE plots}(b) respectively. For both SN and DN phases, the BCP components have maximum density concentrated near the gapped nodal points. Also, there is an anisotropy in the dispersion present through $\gamma k_y\sigma_y$ term which leads to a nonzero transverse TOH response~\cite{Liu,Tanay,Ojasvi}. As seen in Fig~\ref{BCP tensor densityplot}, the BCP components in DN phase are larger in magnitude than SN phase. Consequently, the TOH response becomes higher in DN phase at every value of the Fermi energy than the SN phase, as shown in Fig~\ref{TOHE plots}(a) and Fig~\ref{TOHE plots} (b).

In the NR phase, the Fermi surface appears to be isotropic when $\gamma = 0$. But the system still shows the largest third-order transverse Hall (TOH) response near the band edge, as illustrated in Fig.~\ref{TOHE plots}(d). This occurs because the BCP components in the NR phase contain strongly momentum-dependent anisotropic terms such as $k_x^2$, $k_y^2$ and $k_xk_y$ in the numerator, while the denominator has a ring-like factor $((k^2-k_0^2)^2\lambda^2+\Delta^2)^{5/2}$, which peaks sharply near $k\approx k_0$. As a result, the BCP distribution in momentum space becomes highly localized and anisotropic around the nodal ring. Accordingly, the individual $\chi_{\alpha\beta\gamma\delta}$ components in the transverse conductivity $\chi_{\perp}$ do not cancel each other and give a nonzero response.  Also, in this phase, the system has rotational symmetry, which leads to all the individual TOH conductivity $\chi_{\alpha\beta\gamma\delta}$ components to be nonzero. Consequently, the transverse conductivity $\chi_{\perp}$ does not vanish when $\theta$ becomes integer multiples of $\pi/2$ for all values of Fermi energy, as shown in Fig~\ref{TOHE plots}(c), unlikely the other two phases. We also observe that as we move the Fermi energy far away from the band edge, the TOH responses in every phase start diminishing as the BCP tensor is concentrated near the nodal points. For NR phase the BCP components are sharply localized along the nodal ring only, whereas for SN and DN phases the BCP components are broadly localized near the nodal points. This explains why the TOH response sharply diminishes for NR phase than the other two phases if we move away from the band edges. 

In experiments, the TOH signal may also receive contributions from BCQ~\cite{li2024quantum,Korrapati}, disorder, quantum metric dipole (QMD)~\cite{yu2025quantum}, and quadrupole (QMQ)~\cite{Tong} along with the BCP. Since our system is TRS invariant, BCQ, QMD, and the side-jump contributions are forbidden under symmetry consideration~\cite{Korrapati,Barman}. However, skew-scattering can still generate a third-order Hall signal. This extrinsic mechanism shows a characteristic $\sin{2\theta}$ angular dependency that differs from the BCP induced transverse response~\cite{Barman}. Moreover, the QMQ produces only a longitudinal third-order signal and no transverse component, so it does not overlap with the BCP induced transverse TOH response. 

Here, we discuss a possible experimental prediction of the magnitudes of both SOH and TOH responses across different phases. To obtain realistic predictions, we consider experimentally relevant parameters for two-dimensional nodal Dirac semimetals with a thickness of $t\sim 1~\mathrm{nm}$ configured in the conventional Hall bar geometry with size $l\times w \sim 100\times 20~\mathrm{\mu m}^2$~\cite{Tanay}. Considering a scattering time of $\tau\sim 1\times 10^{-12}~\mathrm{s}$ and a resistivity of $\rho\sim2~\mathrm{\mu\Omega~m}$, a driving current of magnitude $I_0 =0.5~\mathrm{mA}$ can generate an electric field of approximately $\mathrm{E=500~V/cm}$. Using the BCD component $D_{xz}$ values obtained from Fig.~\ref{SOHE Plots}, at $E_F=\mathrm{0.14~eV}$ the estimated SOH conductivities are found to be $\zeta^{\mathrm{SN}}_{yxx}= \mathrm{1.748\times 10^{-12}~A~m~V^{-2}~}$ and $\zeta^{\mathrm{DN}}_{yxx}= \mathrm{5.06\times 10^{-13}~A~m~V^{-2}~}$, yielding SOH voltages ($\mathrm{V_H^{(2)}}\propto \zeta~E^2$) of approximately $\mathrm{V_H^{(2),{SN}}\sim174.8~\mu V}$ and $\mathrm{V_H^{(2),{DN}}\sim50.6~\mu V}$ for the SN and DN phases, respectively. Similarly, when the field angle is around $\theta=\pi/6$ (see Fig.~\ref{TOHE plots} (d)), the estimated TOH conductivities are found to be $\chi_{\perp}^{\mathrm{NR}}=  6.9\times10^{-19}~\mathrm{A\,m^2/V^3}$, $\chi_{\perp}^{\mathrm{DN}}= 2.46\times10^{-19}~\mathrm{A\,m^2/V^3}$, and $\chi_{\perp}^{\mathrm{SN}}= -3.18\times10^{-20}~\mathrm{A\,m^2/V^3}$, corresponding to TOH voltages ($\mathrm{V_H^{(3)}}\propto\chi_{\perp}~E^3$) of approximately $3.45~\mu\mathrm{V}$, $1.23~\mu\mathrm{V}$, and $-0.16~\mu\mathrm{V}$ for the NR, DN, and SN phases, respectively. Although based on a minimal low-energy model with representative parameters, the qualitative features are expected to persist in realistic nodal semimetals. Material-specific predictions for the MX (M = Pd, Pt; X = S, Se, Te) family require effective model parameters obtained from first-principles calculations.


\section{\label{Summary}Summary}
We have studied the nonlinear Hall response in tunable nodal semimetals. By tuning the band parameters, we have traced the transition between single-node, double-node, and nodal-ring phases. Each phase shows a different pattern of quantum geometric quantities and a different hierarchy of nonlinear responses. As the system is TRS invariant and inversion asymmetric, both SN and DN phases exhibit the BCD-induced extrinsic second-order Hall responses. This response in the SN phase turns out to be larger than the DN phase as a manifestation of the distinct distribution of the Berry curvature in the DN phase. For the NR phase, this quantity vanishes as the inversion symmetry is restored. We further find that the BCP becomes finite for all these cases and contributes to a nonzero third-order Hall response. We find that in the NR phase, the BCP components have a larger magnitude and are sharply confined near the band edge along the nodal ring. This leads to an enhanced TOH response however it falls quickly as the Fermi level moves away from the band edge.  It is worth pointing out that the system shows these nonlinear Hall responses inherently in all nodal phases without any tilt or warping of the Fermi surface~\cite{Samal}. In sum, our study highlights 2D semimetallic systems with tunable nodal phases that can be used to control quantum geometry-driven transport phenomena.
  
\section{Acknowledgment}{AD thanks Kush Saha and Snehasish Nandy for useful discussions. AD acknowledges financial support from the Department of Atomic Energy (DAE), Govt. of India, and CSIR for giving fellowship (File No.-09/1002(0049)/2020-EMR-I).}

\appendix

\section*{Appendices}

These appendices provide detailed derivations supporting Eqs.~\ref{TOHcurrentdensity},~\ref{TOHE Conductivity},~\ref{Berry curvature}, and~\ref{BCP components} of the main text. In Appendix~\ref {app: BC_BCP}, we derive the Berry curvature and the Berry connection polarizability for a generic two-band Hamiltonian, and subsequently apply the formalism to the model Hamiltonian discussed in the main text to obtain Eqs.~~\ref{Berry curvature}, and \ref{BCP components}. In Appendix~\ref{sec: TOH_Current_conductivity}, we present the detailed derivation of the third-order Hall current density and conductivity used in Eqs.~\ref{TOHcurrentdensity} and~\ref{TOHE Conductivity} of the main text.

\section{Berry curvature and Berry connection polarizability for two-band Hamiltonian}\label{app: BC_BCP}

We begin with a generic two-band Hamiltonian of the form
\begin{equation}
H(\mathbf{k}) = d_0(\mathbf{k}) + \mathbf{d}(\mathbf{k}) \cdot \boldsymbol{\sigma},
\label{eq: APP_D1}
\end{equation}
where $\mathbf{d}(\mathbf{k}) = (d_x, d_y, d_z)$ and $\boldsymbol{\sigma}$ denotes the Pauli matrices. The corresponding energy eigenvalues are given by 
\begin{equation}
\varepsilon_\pm(\mathbf{k}) = d_0(\mathbf{k}) \pm |\mathbf{d}(\mathbf{k})|.
\label{eq: APP_D2}
\end{equation}
For a generic two-band system with band index $n=\pm$, the Berry curvature can be expressed as~\cite{Di}
\begin{equation}
\Omega^{\pm}_{\alpha\beta}= i \frac{\langle \pm | \partial_\alpha H | \mp \rangle\langle \mp | \partial_\beta H | \pm \rangle- (\alpha \leftrightarrow \beta)}{(\epsilon_{\pm} - \epsilon_{\mp})^2},
\label{eq: APP_D3}
\end{equation}
where $(\alpha,\beta)= (x,y,z)$ are spatial directions, $\partial_{\alpha}=\partial/\partial k_{\alpha}$ and $(\epsilon_{\pm} - \epsilon_{\mp})=\pm 2 |\mathbf{d}|$. The momentum derivative of the Hamiltonian is
\begin{equation}
    \partial_\alpha H = (\partial_\alpha d_0)+(\partial_\alpha d_i)\sigma_i.
    \label{eq: APP_D4}
\end{equation}
Since the scalar term does not contribute between different bands due to the orthogonality of the eigenstates, we obtain
\begin{equation}
    \langle \pm | \partial_\alpha H | \mp \rangle
= (\partial_\alpha d_i)\,\langle \pm | \sigma_i | \mp \rangle.
\label{eq: APP_D5}
\end{equation}
Substituting Eq.~\ref{eq: APP_D5} into Eq.~\ref{eq: APP_D3}, the Berry curvature becomes
\begin{equation}
\Omega^{\pm}_{\alpha\beta}
= \frac{i}{4 |\mathbf{d}|^2}
\left[(\partial_\alpha d_i)(\partial_\beta d_j)
- (\alpha \leftrightarrow \beta)
\right]
\langle \pm | \sigma_i | \mp \rangle
\langle \mp | \sigma_j | \pm \rangle.
\label{eq: APP_D6}
\end{equation}
Using the identity $\sigma_i \sigma_j = \delta_{ij} + i \epsilon_{ijk} \sigma_k$, together with the expectation value $\langle \pm | \sigma_k | \pm \rangle = \pm d_k/|\mathbf{d}|$, we obtain
\begin{equation}
\langle \pm | \sigma_i | \mp \rangle
\langle \mp | \sigma_j | \pm \rangle
= \delta_{ij}
\pm i \epsilon_{ijk} \left(\frac{d_k}{|\mathbf{d}|} \right)
- \frac{d_i d_j}{|\mathbf{d}|^2}.
\label{eq: APP_D7}
\end{equation}
Substituting Eq.~\ref{eq: APP_D7} into Eq.~\ref{eq: APP_D6}, the symmetric terms vanish upon antisymmetrization with respect to $\alpha$ and $\beta$, leaving only the antisymmetric contribution. Consequently, we can write
\begin{equation}
    \Omega^{\pm}_{\alpha\beta}
= \pm \frac{1}{4 |\mathbf{d}|^3}2\epsilon_{ijk}\, d_k\, (\partial_\alpha d_i)\, (\partial_\beta d_j).
\label{eq: APP_D8}
\end{equation}
Finally, the Berry curvature for a generic two-band Hamiltonian can be written compactly in terms of the $\mathbf{d}$ vector as~\cite{Zhuang}
\begin{equation}
    \Omega^{\pm}_{\alpha\beta}(\mathbf{k})= \pm \frac{1}{2}\,
\frac{\mathbf{d}(\mathbf{k}) \cdot\left(\partial_\alpha \mathbf{d}(\mathbf{k})\times\partial_\beta \mathbf{d}(\mathbf{k})\right)}{|\mathbf{d}(\mathbf{k})|^3}.
\label{eq: APP_D9}
\end{equation}

Next, we derive the expression of the Berry connection polarisability, which is defined in Eq.~\ref{eq:BCP} of the main text as
\begin{equation}
 G_{m,\alpha\beta}= 2 \,\mathrm{Re} \sum_{n \ne m}\frac{
A_{mn,\alpha}\, A_{nm,\beta}}{\epsilon_m - \epsilon_n},
\label{eq: APP_D10}
\end{equation}
where the unperturbed Berry connection, $A_{mn,\alpha} = i \langle m | \partial_\alpha n \rangle$. Using the identity $\langle m | \partial_\alpha n \rangle= \langle m | \partial_\alpha H | n \rangle/(\epsilon_n - \epsilon_m)$ for $(m \neq n)$, we can write the Eq.~\ref{eq: APP_D10} for two-band Hamiltonian as
\begin{equation}
G^{\pm}_{\alpha\beta}= 2\,\mathrm{Re} \left[\frac{\langle \pm | \partial_\alpha H | \mp \rangle\langle \mp |\partial_\beta H | \pm \rangle}{(\epsilon_{\pm} - \epsilon_{\mp})^3}\right].
\label{eq: APP_D11}
\end{equation}
Using Eq.~\ref{eq: APP_D5}, this expression can be rewritten as 
\begin{equation}
G^{\pm}_{\alpha\beta}= 2\,\mathrm{Re}\left[\frac{(\partial_\alpha d_i)(\partial_\beta d_j)\langle \pm | \sigma_i | \mp \rangle\langle \mp | \sigma_j | \pm \rangle}{8|\mathbf{d}|^3} \right].
\label{eq: APP_D12}
\end{equation}
Substituting the matrix elements from Eq.~\ref{eq: APP_D7} and taking the real part yields
\begin{equation}
    G^{\pm}_{\alpha\beta}
= \pm \frac{1}{4 |\mathbf{d}|^3}
\left[
(\partial_\alpha d_i)(\partial_\beta d_i)
- \frac{ d_i d_j (\partial_\alpha d_i)
(\partial_\beta d_j)}{|\mathbf{d}|^2}
\right].
\label{eq: APP_D13}
\end{equation}
Thus, the Berry connection polarizability for a generic two-band system can be expressed in terms of the $\mathbf{d}$ vector as~\cite{Zhuang}
\begin{align}
 G^{\pm}_{\alpha\beta}&= \pm \frac{1}{4 |\mathbf{d}|^3}
\left[(\partial_\alpha \mathbf{d}) \cdot (\partial_\beta \mathbf{d})- \frac{( \mathbf{d} \cdot \partial_\alpha \mathbf{d} )( \mathbf{d} \cdot \partial_\beta \mathbf{d} )}{|\mathbf{d}|^2}\right]\nonumber\\
&= \pm \frac{1}{4 |\mathbf{d}(\mathbf{k})|}\, \partial_\alpha\hat{\mathbf{d}}(\mathbf{k})\cdot\partial_\beta \hat{\mathbf{d}}(\mathbf{k}).
\label{eq: APP_D14}
\end{align}

For the model Hamiltonian with a small gap parameter $\Delta$, $\mathbf{d}(\mathbf{k})=(d_x,d_y,d_z)=\bigl(\lambda(k^2-k_0^2),\gamma k_y,\Delta\bigr)$ considered in Eq.~\ref{Hamiltonian} of the main text, the Berry curvature obtained from Eq.~\ref{eq: APP_D9} is
\begin{equation}
    \Omega^{\pm}_{xy}(\mathbf{k})= \pm \frac{\lambda k_x \gamma \, \Delta}{\left[(k^2-k_0^2)^2 \lambda^2 + k_y^2 \gamma^2 + \Delta^2\right]^{3/2}}.
\label{eq: APP_D9_model}
\end{equation}
Since $\Omega^{\pm}_{xy}(\mathbf{k})$ represents the out-of-plane component of the Berry curvature, it may equivalently be denoted as $\Omega^{\pm}_{z}(\mathbf{k})$. This expression corresponds to Eq.~\ref{Berry curvature} of the main text.

Similarly, using Eq.~\ref{eq: APP_D14}, the Berry connection polarizability tensor components for the model Hamiltonian are obtained as 
 \begin{align}
     G_{xx}^{\pm}&=\pm \frac{k_x^2 \left( k_y^2 \gamma^2 + \Delta^2 \right) \lambda^2}
{\left[(k^2-k_0^2)^2 \lambda^2 + k_y^2 \gamma^2 + \Delta^2\right]^{5/2}},\nonumber\\
     G_{yy}^{\pm}&=\pm\frac{\gamma^2 (\Delta^2 + \left( k_0^2 - k_x^2 + k_y^2 \right)^2 \lambda^2 )+4k_y^2 \Delta^2 \lambda^2}
{4\left[(k^2-k_0^2)^2 \lambda^2 + k_y^2 \gamma^2 + \Delta^2\right]^{5/2}}
,\nonumber\\
G_{xy}^{\pm}&= G_{yx}^{\pm}=\pm\frac{k_x k_y (\left( k_0^2 - k_x^2 + k_y^2 \right) \gamma^2 + 2 \Delta^2)\lambda^2}
{2\left[(k^2-k_0^2)^2 \lambda^2 + k_y^2 \gamma^2 + \Delta^2\right]^{5/2}}.
\label{eq: APP_D14_model}
 \end{align}
These expressions reproduce Eq.~\ref{BCP components} of the main text.

\section{Derivation of the third-order Hall current density and conductivity}\label{sec: TOH_Current_conductivity}
In this section, we derive the third-order current density and conductivity [Eq.~\ref{TOHcurrentdensity} and~\ref{TOHE Conductivity} of the main text] starting from the semiclassical expression for the current density~\cite{Liu,Tanay,Ojasvi}, 
\begin{equation}
\mathbf{j}=-\int_k f(\mathbf{k})\dot{\mathbf r}(\mathbf{k}).
\label{eq:s1}
\end{equation}
The semiclassical equation of motion in the presence of an electric field is given by
\begin{equation}
\dot{\mathbf r}=\mathbf{\nabla_k}\epsilon^c-\mathbf{E} \times\mathbf\Omega^c.
\label{eq:s2}
\end{equation}
Here, the higher-order corrections to band energy and Berry curvature due to the applied electric field are expanded as
\begin{equation}
\epsilon^c=\epsilon^{(0)}+\epsilon^{(2)},
\qquad
\mathbf\Omega^c=\mathbf\Omega^{(0)}+\mathbf\Omega^{(1)}.
\label{eq:s3}
\end{equation}
Here, $\epsilon^{(0)}$ and $\mathbf\Omega^{(0)}$ are unperturbed band energy and Berry curvature respectively.
The first-order correction to the band energy is gauge dependent and vanishes within the wave-packet formalism~\cite{Gao,Tanay}. Therefore, the leading correction to the energy appears at second order in the electric field. The nonequilibrium distribution function is obtained from the Boltzmann equation within the relaxation time approximation ( Eq.~\ref{eq:BZrelax} of the main text),
\begin{equation}
\mathbf{\dot{k}}\cdot\mathbf{\nabla_k} f=\frac{f_{eq}-f}{\tau},
\label{eq:s4}
\end{equation}
where $\tau$ is the relaxation time. Defining $D=\mathbf{\dot{k}}\cdot\mathbf{\nabla_k}= -\mathbf{E}\cdot\mathbf{\nabla_k}$, the iterative solution becomes
\begin{equation}
f=\sum_{i=0}^{\infty}(\tau D)^i f_{eq}(\epsilon^c).
\label{eq:s5}
\end{equation}
To obtain the third-order current response, we expand the distribution function up to cubic order in the electric field,
\begin{equation}
f=f_{eq}+\tau Df_{eq}+\tau^2D^2f_{eq}+\tau^3D^3f_{eq}.
\label{eq:s6}
\end{equation}
Since the equilibrium distribution depends on the corrected energy, we Taylor expand $f_{eq}(\epsilon^c)$ around $\epsilon^{(0)}$ as
\begin{equation}
f_{eq}(\epsilon^c)=f_{eq}(\epsilon^{(0)})+\epsilon^{(2)}f_{eq}'+.....,
\label{eq:s7}
\end{equation}
where $f_{eq}'=\frac{\partial f_{eq}}{\partial \epsilon}|_{\epsilon=\epsilon^{(0)}}$. Substituting Eq.~\ref{eq:s7} into Eq.~\ref{eq:s6}, we obtain
\begin{align}
f=&f_{eq}+\epsilon^{(2)}f_{eq}'+\tau Df_{eq}+\tau D(\epsilon^{(2)}f_{eq}')+\tau^2D^2f_{eq}\nonumber\\
&+\tau^3D^3f_{eq}.
\label{eq:s8}
\end{align}
Substituting Eqs.~\ref{eq:s2} and \ref{eq:s8} into the current density expression in Eq.~\ref{eq:s1} and retaining only the cubic-order terms in the electric field, the third-order current density becomes
\begin{widetext}
\begin{align}
j^{(3)}
=&
\int_k
(\mathbf E\times\mathbf\Omega^{(0)})
\epsilon^{(2)}f_{eq}'
-\tau\int_k
\nabla_k\epsilon^{(0)}
(\mathbf E\cdot\nabla_k)\epsilon^{(2)}f'_{eq}-\tau\int_k
\nabla_k\epsilon^{(2)}
(\mathbf E\cdot\nabla_k)f_{eq}+\tau\int_k
(\mathbf E\times\mathbf\Omega^{(1)})
(\mathbf E\cdot\nabla_k)f_{eq}
\nonumber\\
&
+\tau^2\int_k
(\mathbf E\times\mathbf\Omega^{(0)})
(\mathbf E\cdot\nabla_k)^2f_{eq}
-\tau^3\int_k
\nabla_k\epsilon^{(0)}
(\mathbf E\cdot\nabla_k)^3f_{eq} .
\label{eq:s9}
\end{align}
\end{widetext}
This is the same expression as the third-order current density presented in Eq.~\ref{TOHcurrentdensity} of the main text.

Since the unperturbed Berry curvature $\mathbf{\Omega}^{(0)}$ is odd under time-reversal symmetry, the integrands of the first and fifth terms are odd in momentum space and therefore vanish upon momentum integration. Hence, the surviving terms are
\begin{widetext}
\begin{align}
j^{(3)}
=&
-\tau\int_k
\nabla_k\epsilon^{(0)}
(\mathbf E\cdot\nabla_k)\epsilon^{(2)}f'_{eq}-\tau\int_k
\nabla_k\epsilon^{(2)}
(\mathbf E\cdot\nabla_k)f_{eq}+\tau\int_k
(\mathbf E\times\mathbf\Omega^{(1)})
(\mathbf E\cdot\nabla_k)f_{eq}
-\tau^3\int_k
\nabla_k\epsilon^{(0)}
(\mathbf E\cdot\nabla_k)^3f_{eq} .
\label{eq:s10}
\end{align}    
\end{widetext}
The above expression can also be expressed in terms of third-order conductivity $\chi$ as 
\begin{equation}    j_{\alpha}^{(3)}=\chi_{\alpha\beta\gamma\delta}E_{\beta}E_{\gamma}E_{\delta},
\label{eq:s11}
\end{equation}
where the indices $\alpha, \beta, \gamma, \delta \in \{x,y\}.$
This fourth-rank conductivity tensor contains two distinct contributions,
\begin{equation}
\chi_{\alpha\beta\gamma\delta}=\chi^{I}_{\alpha\beta\gamma\delta}+\chi^{II}_{\alpha\beta\gamma\delta},
\label{eq:s12}
\end{equation}
where the first contribution $\chi^{I}_{\alpha\beta\gamma\delta}$ is proportional to the relaxation time $\tau$, whereas the second contribution  $\chi^{II}_{\alpha\beta\gamma\delta}$ varies as $\tau^3$. Using the band velocity $v_{\alpha}=\partial_{k_a}\epsilon^{(0)}$, the $\tau^3$ contribution to the third-order conductivity arises from the last term of Eq.~\ref{eq:s10} can be written as 
\begin{equation}
j_{\alpha,II}^{(3)}= -\tau^3
\int_k v_{\alpha} (\mathbf E\cdot\nabla_k)^3 f_{eq}.
\label{eq:s13}
\end{equation}
Expanding the cubic differential operator as $(\mathbf E\cdot\nabla_k)^3=(E_b\partial_b)(E_c\partial_c)(E_d\partial_d)$, and substituting this into the current expression and comparing with Eq.~\ref{eq:s11}, the $\tau^3$ contribution to the third-order conductivity tensor is obtained as
\begin{equation}
\chi^{II}_{abcd}=-\tau^3
\int_k\,v_{\alpha}
\partial_\beta\partial_\gamma\partial_\delta f_{eq}.
\label{eq:s14}
\end{equation}
The contribution linear in relaxation time $\tau$ arises from the remaining three terms in Eq.~\ref{eq:s10}. Thus, the corresponding third-order current density can be written as
\begin{align}
j_{\alpha,I}^{(3)}
&=-\tau \int_{\mathbf{k}}v_{\alpha}\,(\mathbf{E}\!\cdot\!\nabla_{\mathbf{k}})\epsilon^{(2)}
\,f_{eq}'-\tau \int_{\mathbf{k}}
\left(\partial_{\alpha}\epsilon^{(2)}\right)
(\mathbf{E}\!\cdot\!\nabla_{\mathbf{k}})
f_{eq}\nonumber\\&+\tau\int_{\mathbf{k}}\left(\mathbf{E}\times\boldsymbol{\Omega}^{(1)}\right)_{\alpha}
(\mathbf{E}\!\cdot\!\nabla_{\mathbf{k}})f_{eq}.
\label{eq:s15}
\end{align}
The first term in the RHS of the above equation can be rewritten as 
\begin{equation}
T_1=-\tau\int_k v_{\alpha}E_{\beta}\left(\partial_{\beta}\epsilon^{(2)}\right) f_{eq}'-\tau\int_k v_{\alpha}E_{\beta}\epsilon^{(2)}\left(\partial_{\beta}f_{eq}'\right) 
\label{eq:s16}
\end{equation}
In the main text, we have already shown that the second-order correction to the band energy can be expressed in terms of the Berry connection polarizability tensor as
\begin{equation}
\epsilon^{(2)}=\frac{1}{2}G_{\gamma\delta}\,E_{\gamma}E_{\delta}.
\label{eq:s17}
\end{equation}
Substituting Eq.~\ref{eq:s17} into Eq.~\ref{eq:s16}, the first term becomes
\begin{align}
T_1=&-\frac{\tau}{2}E_{\beta}E_{\gamma}E_{\delta}\int_k \left[v_{\alpha}(\partial_{\beta}G_{\gamma\delta})f_{eq}'- v_{\alpha}G_{\gamma\delta}(\partial_{\beta}f_{eq}')\right].
\label{eq:s18}
\end{align}
Using the relations $v_{\alpha}f_{eq}'=\partial_{\alpha}f_{eq}$, $\partial_{\beta}f_{eq}'=v_{\beta} f_{eq}''$, and performing the integration by parts in the first term of $T_1$, we obtain
\begin{align}
T_1=\frac{\tau}{2}E_{\beta}E_{\gamma}E_{\delta}\int_k
\left[f_{eq}(\partial_{\alpha}\partial_{\beta}G_{\gamma\delta})-v_{\alpha}v_{\beta}G_{\gamma\delta}f_{eq}''\right].
\label{eq:s19}
\end{align}
Next, the second term in Eq.~\ref{eq:s15} becomes
\begin{align}
T_2
=&
-\tau
\int_k
(\partial_{\alpha}\epsilon^{(2)})
(\mathbf E\cdot\nabla_k)f_{eq}
\nonumber\\
=&-\frac{\tau}{2}E_{\beta}E_{\gamma}E_{\delta}\int_k(\partial_{\alpha}G_{\gamma\delta})(\partial_{\beta}f_{eq}).
\label{eq:s20}
\end{align}
Again, applying integration by parts leads to
\begin{equation}
T_2=\frac{\tau}{2}E_{\beta}E_{\gamma}E_{\delta}\int_k
f_{eq}(\partial_{\beta}\partial_{\alpha}G_{\gamma\delta}).
\label{eq:s21}
\end{equation}
To evaluate the third term in Eq.~\ref{eq:s15}, we use the first-order correction to the Berry connection, which we have also derived in the main text as $A_{\mu}^{(1)}=G_{\mu\nu}E_{\nu}$, which gives the first-order correction to the Berry curvature as
\begin{equation}
\Omega_{\lambda}^{(1)}=\epsilon_{\lambda\mu\nu}
(\partial_{\mu}G_{\nu\gamma})E_{\gamma}.
\label{eq:s22} 
\end{equation}
Therefore, 
\begin{align}
(\mathbf E\times\mathbf\Omega^{(1)})_{\alpha}
=&\epsilon_{\alpha\beta\lambda}E_{\beta}\Omega_{\lambda}^{(1)}
\nonumber\\=&\epsilon_{\alpha\beta\lambda}\epsilon_{\lambda\mu\nu}E_{\beta}(\partial_{\mu}G_{\nu\gamma})E_{\gamma}.
\label{eq:s23}
\end{align}
Using the identity $\epsilon_{\alpha\beta\lambda}
\epsilon_{\lambda\mu\nu}=\delta_{\alpha\mu}\delta_{\beta\nu}-
\delta_{\alpha\nu}\delta_{\beta\mu}$, we obtain
\begin{equation}
(\mathbf E\times\mathbf\Omega^{(1)})_{\alpha}=E_{\beta}E_{\gamma}
\left(\partial_{\alpha}G_{\beta\gamma}-\partial_{\beta}G_{\alpha\gamma}\right).
\label{eq:s24}
\end{equation}
Substituting Eq.~\ref{eq:s24} into the third term of Eq.~\ref{eq:s15}, we obtain
\begin{align}
T_3&=\tau\int_{\mathbf{k}}\left(\mathbf{E}\times\boldsymbol{\Omega}^{(1)}\right)_{\alpha}(\mathbf{E}\!\cdot\!\nabla_{\mathbf{k}})f_{eq}\nonumber\\&= \tau
E_{\beta}E_{\gamma}E_{\delta}\int_k\left(\partial_{\alpha}G_{\beta\gamma}-\partial_{\beta}G_{\alpha\gamma}\right)(\partial_{\delta}f_{eq}).
\label{eq:s25}
\end{align}
Performing integration by parts yields
\begin{equation}
T_3 =-\tau E_{\beta}E_{\gamma}E_{\delta}\int_k
f_{eq}\left(\partial_{\alpha}\partial_{\delta}G_{\beta\gamma}-\partial_{\beta}\partial_{\delta}G_{\alpha\gamma}\right).
\label{eq:s26}
\end{equation}
Finally, combining Eqs.~\ref{eq:s19}, \ref{eq:s21}, and \ref{eq:s26}, the contribution to the third-order conductivity tensor linear in $\tau$ becomes
\begin{align}
\chi^{I}{\alpha\beta\gamma\delta}=&\tau\int_k \Big(
\partial_{\alpha}\partial_{\beta}G_{\gamma\delta}-
\partial_{\alpha}\partial_{\delta}G_{\beta\gamma}+
\partial_{\beta}\partial_{\delta}G_{\alpha\gamma} \Big)f_{eq}
\nonumber\\&-\frac{\tau}{2}\int_kv_{\alpha}v_{\beta}G_{\gamma\delta}f_{eq}''.
\label{eq:s31}
\end{align}
This corresponds to Eq.~\ref{TOHE Conductivity} of the main text.

\bibliography{references}


	
\end{document}